\documentclass[a4paper,11pt]{article}
\usepackage[a4paper,margin=2.35cm]{geometry}
\usepackage{amsmath,amssymb,bm,mathtools,booktabs,graphicx,subcaption,float}
\usepackage{hyperref}
\usepackage{titlesec}
\usepackage{xcolor}
\usepackage{listings}
\usepackage{indentfirst}

\hypersetup{colorlinks=true,linkcolor=blue,citecolor=blue,urlcolor=blue}
\lstdefinelanguage{Matlab}{morekeywords={break,case,catch,continue,else,elseif,end,for,function,global,if,otherwise,persistent,return,switch,try,while},sensitive=true,morecomment=[l]\%,morestring=[m]'}
\newcommand{\Tr}{\operatorname{Tr}}
\newcommand{\Id}{\mathbb I}

\newcommand{\ket}[1]{\left|#1\right\rangle}
\newcommand{\bra}[1]{\left\langle#1\right|}

\newcommand{\calS}{\mathcal S}

\renewcommand{\thesection}{\Roman{section}}
\titleformat{\subsection}{\normalfont\large\bfseries\centering}{\thesubsection}{1em}{}
\renewcommand{\thesubsection}{\Alph{subsection}}
\renewcommand{\thesubsubsection}{\alph{subsubsection}}
\title{\textbf{Nonlocality of Tree-Structured Quantum Networks under Relativistic Motion}}
\author{Tianci Feng, Tinggui Zhang$^\dagger$\\
	\small School of Mathematics and Statistics, Hainan Normal University, Haikou, 571158, China\\
	\small $^\dagger$ Correspondence to: \href{mailto:tinggui333@163.com}{tinggui333@163.com}}           
\date{}  %

\begin{document}
\maketitle
\begin{abstract}
We investigate how relativistic acceleration modifies source-independent nonlocality in tree-structured quantum networks within a weak-coupling Unruh--DeWitt detector model. We consider each edge affected by acceleration and show that, when local product measurements are performed independently at each node, the attainable upper bound of the network nonlocality witness is determined by the two largest singular values of the correlation matrix of each edge. For a seven-node binary tree with maximally entangled inputs, the decay of the nonlocality witness is governed entirely by the number of edges perturbed by acceleration. After extending the conclusions to $n$-layer complete binary trees, the governing topological scaling variable is the ratio of perturbed edges to leaf nodes. The network depth gives rise to two opposite behaviors: if only the root node is accelerated, the number of perturbed edges remains unchanged no matter how the network expands in scale, so the influence brought by local acceleration is relatively weakened and the nonlocality witness shows almost no obvious change with depth; if all leaf nodes are accelerated simultaneously, the nonlocality witness is exactly the same at different depths. We also calculate the Werner-state visibility threshold required by this nonlocality witness. This work clarifies how network topology determines the robustness of nonlocality under relativistic perturbations, and clearly delineates the range of applicability of the effective-state model.
\end{abstract}

\section{Introduction}
Quantum networks consist of multiple entanglement sources, local measurement nodes, and quantum channels connecting those nodes. Unlike the standard Bell scenario, a network-local model requires not only that each output be determined by a local measurement input and the hidden variables carried by adjacent sources, but also that the hidden variables emitted by distinct sources be statistically independent. Source independence therefore imposes nonlinear constraints on the joint distribution and leads to network criteria that differ from standard Bell locality, including bilocal, star-shaped, chain, and tree-shaped scenarios \cite{Brunner2014,Branciard2010,Branciard2012,Tavakoli2014,Tavakoli2022,Luo2018}. Network nonlocality is consequently determined not only by the entanglement quality of individual edges, but also by the way independent sources are assembled according to the network topology.

A range of network-nonlocality criteria has been developed for different architectures. Branciard and collaborators characterized bilocal correlations in entanglement-swapping experiments, establishing a basis for testing source independence \cite{Branciard2010,Branciard2012}. Star networks subsequently revealed how the central-node degree, the number of leaves, and the number of independent sources affect $n$-local inequalities \cite{Tavakoli2014,Andreoli2017,Kundu2020}; independent-source nonlocality has also been tested experimentally in star, chain, and bilocal configurations \cite{Carvacho2016,Sun2019}. Tree networks extend these ideas to multilayer acyclic structures. Yang et al. studied nonlocal correlations in two-forked tree-tensor networks \cite{Yang2021}, and later considered forked tree-shaped networks with arbitrary branching and their robustness to noise \cite{Yang2022Tree}. These results show that tree topologies provide a natural setting in which the number of nodes, edges, branches, and layers can be varied while source independence is retained. Recent work has further extended network nonlocality to full network nonlocality, network self-testing, and scenarios with weaker source-independence assumptions \cite{Supic2020,Supic2021,Renou2019,Pozas2022}.

A complete binary tree is a particularly transparent member of this class. Its $k$th layer contains $2^{k-1}$ nodes, so the total number of nodes, independent sources, and leaves obey fixed exponential scaling laws with the depth $n$. Unlike a chain, where the main structural variable is the propagation length, or a star, where the central degree is the dominant variable, a complete binary tree permits the depth, width, and number of leaves to be varied systematically while keeping local degrees bounded. It is precisely this structure that makes it possible to quantitatively study the accumulation and attenuation effects of a fixed local perturbation as the network expands in size.

Experimental progress on independent-source networks has also connected network criteria with practical multi-node quantum platforms \cite{Pompili2021,Gu2023}. Relativistic motion introduces a further class of local perturbations. The interaction between an accelerated detector and a quantum field changes the local detector state and the spectrum of its correlation matrix. The Unruh--DeWitt (UDW) model provides a standard framework for describing the coupling between a localized two-level system and a quantum field \cite{DeWitt1979,BirrellDavies,Landulfo2009,Salton2015,Grimmer2021}. Existing work on relativistic quantum networks has mainly addressed chain and star configurations \cite{Li2026}. In multilayer tree networks, however, it remains unclear how a local acceleration propagates through the singular values of the affected edge resources, how this propagation depends on depth, and how a change in singular-value ordering should be interpreted in an extrapolated strong-acceleration regime.

This work asks how network depth, the number of affected edges, and the location of accelerated nodes jointly determine the decay scaling of a source-independent nonlocality witness under a fixed local acceleration model and measurement strategy. Rather than solving a complete field-theoretic master equation, we construct effective edge states within a weak-coupling, first-order perturbative framework and insert the singular values of their correlation matrices into a tree-network criterion. This produces a unified route from local detector dynamics to network-scale behavior: the UDW edge states are first expressed in an edge-wise singular-value representation; single-node and nonadjacent double-node acceleration are then compared in the seven-node network to isolate the role of the affected-edge number; finally, the analysis is extended to an $n$-layer complete binary tree to compare root-only and all-leaf acceleration and to derive the Werner visibility threshold. We also numerically check the singular-value crossing at $\eta=2$, distinguish the controlled perturbative branch from the strong-acceleration extrapolation, and use a trace-preservation test to state the validity boundary of the effective-state model. The existing network-locality criterion is kept unchanged; the contribution is to determine how a relativistic local perturbation is transmitted to the network scale through the spectrum of the edge resources.

The scope of the analysis is restricted to independent edge environments, maximally entangled inputs, at most one accelerated endpoint per edge, and the internal-node product-measurement strategy adopted here. Sec.~II defines the detector and source-independent-network models; Sec.~III derives the edge states for the two single-endpoint acceleration protocols; Sec.~IV establishes the seven-node witness and its measurement construction; Sec.~V analyzes acceleration location and affected-edge number; and Sec.~VI extends the results to $n$-layer complete binary trees and Werner thresholds. The appendices contain the field-theory expansion, the acceleration of detector $E$, spectral-crossing checks, the trace-preservation test, and numerical reproducibility details. Throughout, “tree-structured quantum network” denotes a source-independent distribution network and not a tensor-network variational algorithm.

\section{Theoretical Model and Network Criterion}

\subsection{Relativistic Detector Dynamics and Effective Edge States}

We use the Unruh--DeWitt detector model to describe the interaction between a locally accelerated quantum system and an external quantum field. Each qubit is represented by a two-level detector with ground and excited states $\ket{0}$ and $\ket{1}$, separated by an energy gap $\Omega$.

Consider an arbitrary edge of the tree, with endpoint detectors $D$ and $E$. An independent entanglement source distributes an entangled detector pair to the two endpoints. One detector is then uniformly accelerated during a finite proper-time interval $[\tau_0,\tau_0+\Delta]$ and coupled to a massless scalar field $\phi$, while the other remains inertial and uncoupled. After the interaction, tracing out the field yields the reduced two-detector state.

The central worldline of the accelerated detector is
\begin{equation}
	t(\tau)
	=
	a^{-1}\sinh(a\tau),
	\qquad
	x(\tau)
	=
	a^{-1}\cosh(a\tau),
	\qquad
	y(\tau)=z(\tau)=0.
\label{orig:1}\end{equation}
where $a$ is the proper acceleration and $\tau$ is the detector proper time.

To preserve source independence after the local evolution, we assume independent environments for different edges and neglect field-mediated cross-correlations between edges. Each edge therefore evolves separately, and the network resource remains a tensor product of edge states.

We use natural units
	$c=\hbar=k_{\mathrm B}=1.$
The free detector Hamiltonians are
\begin{equation}
	H_D=\Omega D^\dagger D,
	\qquad
	H_E=\Omega E^\dagger E.
\label{orig:2}\end{equation}

Here $D$ and $D^\dagger$ are the lowering and raising operators of detector $D$, satisfying
\begin{equation}
	D\ket{1}=\ket{0},
	\qquad
	D^\dagger\ket{0}=\ket{1},
	\qquad
	D\ket{0}=D^\dagger\ket{1}=0.
\label{orig:3}\end{equation}
Detector $E$ obeys the same relations.

The initial two-detector state is
\begin{equation}
	\ket{\psi_\theta}_{DE}
	=
	\sin\theta\,\ket{01}
	+
	\cos\theta\,\ket{10},
	\qquad
	0\leq\theta\leq\frac{\pi}{2}.
\label{orig:4}\end{equation}

For $\theta=\pi/4$ the state is maximally entangled; for $\theta=0$ or $\theta=\pi/2$ it is a product state.

The massless scalar field is initially in the Minkowski vacuum $\ket{0_M}$ and is uncorrelated with the detectors. The joint initial state is therefore
\begin{equation}
	\ket{\Psi_{\tau_0}^{DE\phi}}
	=
	\ket{\psi_\theta}_{DE}
	\otimes
	\ket{0_M}_{\phi}.
\label{orig:5}\end{equation}

Let $H_{\mathrm{KG}}$ denote the free Hamiltonian of the massless scalar field. The total detector--field Hamiltonian is
\begin{equation}
	H_{DE\phi}
	=
	H_D+H_E+H_{\mathrm{KG}}+H_\mathrm{int}.
\label{orig:6}\end{equation}
Here $H_{\mathrm{KG}}$ is the free field Hamiltonian and $H_{\mathrm{int}}$ describes detector--field coupling, with $H_{\mathrm{int}}=H_{\mathrm{int}}^D+H_{\mathrm{int}}^E$. In a single-endpoint protocol only the coupling of the selected endpoint is switched on.
	
In the interaction picture, the single-endpoint coupling is given below. To keep the main text focused, we list only the compact evolution operator and the Rindler--Minkowski mode relation needed to construct the effective edge state; the spatial smearing, Dyson expansion, and mode-function derivations are given in Appendices A.1--A.3.
\begin{equation}
	\begin{aligned}
		H_I(\tau) &= \lambda_{\mathrm c}g(\tau) &\left[ e^{-\mathrm{i}\Omega\tau} D \otimes \Phi_\chi(\tau) + e^{\mathrm{i}\Omega\tau} D^\dagger \otimes \Phi_\chi^\dagger(\tau) \right].
	\end{aligned}
\label{orig:9}\end{equation}
\begin{equation}
	U^{(1)} = \mathbb{I} - \mathrm{i}\lambda_{\mathrm c} \big[ D\phi(f) + D^\dagger \phi^\dagger(f) \big].
\label{orig:14}\end{equation}
\begin{equation}
	a_{\mathrm{RI}}(\bar{\lambda_{\mathrm f}})
	=
	\frac{
		a_M(F_{1\Omega})
		+
		\mathrm{e}^{-\pi\Omega/a}
		a_M^\dagger(F_{2\Omega})
	}{
		\sqrt{1-\mathrm{e}^{-2\pi\Omega/a}}
	},
\label{orig:16}\end{equation}
Here $\lambda_{\mathrm c}$ is the detector--field coupling constant, $g(\tau)$ is a finite-time switching function, and $f$ contains the transition phase and spatial profile. We keep $\lambda_{\mathrm c}$, $\lambda_e$, and $\lambda_{\mathrm f}$ distinct: they denote the coupling constant, the hidden variable emitted by source $e$, and the field-theory test function, respectively. The proper acceleration is denoted by $a$, whereas the root output in the network model is denoted by $o_A$.

For reproducibility, we use the compact smooth switching profile
\begin{equation}
g_{\Delta,\sigma}(\tau)=h_{\sigma}(\tau-\tau_0)h_{\sigma}(\tau_0+\Delta-\tau),\qquad 0<\sigma<\Delta/2,
\label{new:switch}\end{equation}
where
\begin{equation*}
h_{\sigma}(x)=\begin{cases}0,&x\leq0,\\
\tfrac12[1-\cos(\pi x/\sigma)],&0<x<\sigma,\\
1,&x\geq\sigma.\end{cases}
\end{equation*}
Thus $g_{\Delta,\sigma}$ has compact support in $[\tau_0,\tau_0+\Delta]$ and smooth edges of width $\sigma$. We use the normalized Gaussian profile $\chi_{\kappa}(\boldsymbol\xi)=(2\pi\kappa^2)^{-3/2}\exp[-|\boldsymbol\xi|^2/(2\kappa^2)]$. Minkowski modes are normalized as $u_{\boldsymbol k}(x)=[(2\pi)^3 2\omega_{\boldsymbol k}]^{-1/2}e^{-i\omega_{\boldsymbol k}t+i\boldsymbol k\cdot\boldsymbol x}$, with $[a_{\boldsymbol k},a^\dagger_{\boldsymbol k'}]=\delta^{(3)}(\boldsymbol k-\boldsymbol k')$, and the Rindler wave packets satisfy $||\widetilde F_i||=1$.

Let $\rho_{DE}^{(0)}=|\psi_\theta\rangle_{DE}\langle\psi_\theta|$. The state used below is defined by tracing out the field and applying one global normalization:
\begin{equation}
\widetilde\rho_{DE}=\frac{\Tr_\phi[U^{(1)}(\rho_{DE}^{(0)}\otimes|0_M\rangle\langle0_M|)U^{(1)\dagger}]}{\Tr[U^{(1)}(\rho_{DE}^{(0)}\otimes|0_M\rangle\langle0_M|)U^{(1)\dagger}]}.
\label{new:effective}\end{equation}
We call \eqref{new:effective}, and the equivalent X state in Eq. \eqref{orig:28}, the phenomenological effective resource state (normalized truncated state). It is the resource model used in the network calculation, not a complete local CPTP-channel derivation.

Introduce $q=e^{-2\pi\Omega/a}$ with $0\le q<1$. At fixed $\Omega$, $q$ increases with $a$; $q\to0$ is the zero-acceleration limit and $q\to1$ is only a formal infinite-acceleration limit.

For a finite interaction, the first-order response is controlled by the smeared Wightman function
\begin{equation}
\mathcal{F}_{g,\chi}(\Omega,a)=\int d\tau\,d\tau'\,g(\tau)g(\tau')e^{-i\Omega(\tau-\tau')}W_{\chi,a}(\tau,\tau'),
\label{new:finite_response}\end{equation}
where $W_{\chi,a}$ is the Minkowski-vacuum two-point function along the accelerated trajectory, smeared with $\chi_{\kappa}$. Hence a finite-time response depends on $g$, $\kappa$, $a$, and trajectory details, rather than on $q$ alone. We use the controlled long-time, near-resonance, single-mode approximation $\Omega^{-1}\ll\Delta$, $\sigma\ll\Delta$, $a\kappa\ll1$, and $\Omega\kappa\lesssim1$:
\begin{equation}
\nu^2=\|\lambda_{\mathrm f}\|^2=\lambda_{\mathrm c}^2\frac{\Omega\Delta}{2\pi}e^{-\Omega^2\kappa^2},\qquad \Omega^{-1}\ll\Delta.
\label{orig:19}\end{equation}
Equation \eqref{orig:19} is a narrow-band approximation to \eqref{new:finite_response}, not an exact formula for arbitrary switching. When the conditions fail, use \eqref{new:finite_response} directly and replace $\nu^2$ by $\lambda_{\mathrm c}^2\mathcal{F}_{g,\chi}$. The first-order approximation requires $\nu^2\ll1$, while the high-acceleration control parameter is
\begin{equation}
\eta(q)=\frac{\nu^2(1+q)}{2(1-q)}\ll1.
\label{orig:20}\end{equation}
As $q\to1$, $\eta(q)$ grows rapidly, so a small bare coupling does not control the strong-acceleration extrapolation.

Appendices A.1--A.3 give the complete definitions of the smearing, switching, Dyson expansion, Bogoliubov transformation, and wave packets; the corresponding evolution for detector $E$ is given in Appendix A.4.

\subsection{Source-Independent Network Model}

Let $\mathcal V$ be the set of nodes and $\mathcal E$ the set of independent sources. Each source $e\in\mathcal E$ carries a hidden variable $\lambda_e$, while node $v$ responds only to its measurement input $x_v$ and the hidden variables incident on $v$. The network-local model is
\begin{equation}
P(\bm o|\bm x)=
\int\prod_{e\in\mathcal E}d\mu_e(\lambda_e)
\prod_{v\in\mathcal V}P_v(o_v|x_v,\bm\lambda_v),
\qquad
\int d\mu_e(\lambda_e)=1.
\label{orig:21}\end{equation}
The product structure expresses source independence. A distribution that admits only a single global hidden-variable Bell decomposition need not satisfy the corresponding $m$-local model written in Eq. \eqref{orig:21}.

For binary inputs and outputs, let $A^v_{x_v}$ take values $\pm1$ and define the correlators
\begin{equation}
\left\langle\prod_{v\in\mathcal V}A^v_{x_v}\right\rangle
=\sum_{\bm o}(-1)^{\sum_v o_v}P(\bm o|\bm x).
\label{orig:22}\end{equation}
The nonlinear network inequalities average correlators over different measurement inputs and combine source independence with the network topology \cite{Tavakoli2022,Luo2018}.

\subsection{Two-Qubit States and the Correlation Matrix}

Any two-qubit state can be written in Bloch--Fano form
\begin{equation}
\rho=\frac14\left[
\Id\otimes\Id+\bm r\cdot\bm\sigma\otimes\Id+
\Id\otimes\bm s\cdot\bm\sigma+
\sum_{\mu,\nu=x,y,z}t_{\mu\nu}\sigma_\mu\otimes\sigma_\nu
\right],
\label{orig:23}\end{equation}
where $T=(t_{\mu\nu})$ is the correlation matrix and $t_{\mu\nu}=\Tr(\rho\,\sigma_\mu\otimes\sigma_\nu)$.
\begin{equation}
R=\sqrt{T^\dagger T},\qquad
\tau_1\geq\tau_2\geq\tau_3\geq0
\label{orig:24}\end{equation}
The quantities $\tau_i$ are the eigenvalues of $R$, equivalently the singular values of $T$. We use this notation throughout; subsequent product formulas are written directly in terms of singular values. Once the two-qubit state and its correlation spectrum are fixed, the effective edge state generated by relativistic motion can be inserted into the network calculation.

\section{Relativistic Edge-State Dynamics under Single-Endpoint Acceleration}
\subsection{Acceleration of Detector $D$ and the Rindler-Horizon Effect}

Starting from the initial state, the weak-coupling evolution is obtained by inserting the interaction Hamiltonian in Eq. \eqref{orig:9} into the Rindler expansion and tracing out the field. Retaining the first-order term gives the final state
\begin{equation}
	|\Psi_{\mathrm f}^{DE\phi}\rangle = U^{(1)}|\Psi_{\tau_0}^{DE\phi}\rangle \approx |\Psi_{\tau_0}^{DE\phi}\rangle -i
	\lambda_{\mathrm c}\big[\phi(f)D + \phi^\dagger(f)D^\dagger\big]
	\big(\sin\theta|01\rangle + \cos\theta|10\rangle\big)|0_M\rangle. 
\label{orig:25}\end{equation}

Use the basis convention $|10\rangle\equiv|1_D0_E\rangle$ and $|01\rangle\equiv|0_D1_E\rangle$. The operator relations $D|01\rangle=0$, $D|10\rangle=|00\rangle$, and $D^\dagger|01\rangle=|11\rangle$ show that the interaction produces two transitions: de-excitation $|10\rangle\to|00\rangle$ and excitation $|01\rangle\to|11\rangle$.
Expanding the field operator in Rindler wedge I and projecting onto the relevant mode, Eq. \eqref{orig:25} becomes
\begin{equation}
	|\Psi_{\mathrm f}^{DE\phi}\rangle = |\Psi_{\tau_0}^{DE\phi}\rangle + \frac{\nu}{\sqrt{1-q}}\left[\cos\theta|00\rangle|1_{\widetilde F_1}\rangle + e^{-\pi\Omega/a}\sin\theta|11\rangle|1_{\widetilde F_2}\rangle\right], 
\label{orig:26}\end{equation}

The states $|1_{\widetilde F_i}\rangle$ are normalized one-particle Minkowski wave-packet states, with $\widetilde F_i=F_i/\nu$. We impose orthogonality as a modeling condition, $\langle1_{\widetilde F_1}|1_{\widetilde F_2}\rangle=0$ (equivalently $(\widetilde F_1,\widetilde F_2)_{\rm KG}=0$). Finite switching and spatial smearing do not by themselves guarantee this condition; if an overlap $\epsilon\ne0$ is retained, the reduced state acquires field-induced cross terms and a modified normalization. The first-order joint state incorporates the thermal structure of the Minkowski vacuum through $e^{-\pi\Omega/a}$. With the Unruh temperature $T_U=a/(2\pi)$, this factor is $e^{-\Omega/(2T_U)}=\sqrt q$, while $q=e^{-\Omega/T_U}$ is the corresponding thermal weight. Within the present effective-state model, the acceleration parameter changes the populations and the off-diagonal coherence of the detector state.

Because the final state still contains field degrees of freedom and is not normalized, we take its norm and trace out the field to obtain the reduced detector state $\rho_{DE}$. The normalization factor is
\begin{equation}
	\mathcal{N}_L :=\left\|\Psi_{\mathrm f}^{DE\phi}\right\|^2= 1 + \frac{\nu^2\left(q\sin^2\theta + \cos^2\theta\right)}{1-q}
\label{orig:27}\end{equation}
Here $q\equiv e^{-2\pi\Omega/a}$ is the dimensionless Unruh parameter and $\nu$ is the first-order coupling amplitude.

After tracing out the field, the detector state for acceleration of $D$ takes, in the basis $\{|00\rangle,|01\rangle,|10\rangle,|11\rangle\}$, the standard X form

\begin{equation}
\rho_{\mathrm f}^{(L)}=\mathcal N_L^{-1}
\operatorname{Tr}_{\phi}\left(\ket{\Psi_{\mathrm f}^{DE\phi}}\bra{\Psi_{\mathrm f}^{DE\phi}}\right)
=\begin{pmatrix}
\gamma_L&0&0&0\\
0&2\alpha_L\sin^2\theta&\alpha_L\sin2\theta&0\\
0&\alpha_L\sin2\theta&2\alpha_L\cos^2\theta&0\\
0&0&0&\beta_L
\end{pmatrix},
\label{orig:28}\end{equation}
where
\begin{align}
\alpha_L&=\frac{1-q}{2(1-q)+2\nu^2(q\sin^2\theta+\cos^2\theta)},\label{orig:29}\\
\beta_L&=\frac{\nu^2q\sin^2\theta}{(1-q)+\nu^2(q\sin^2\theta+\cos^2\theta)},\label{orig:30}\\
\gamma_L&=\frac{\nu^2\cos^2\theta}{(1-q)+\nu^2(q\sin^2\theta+\cos^2\theta)}.
\label{orig:31}\end{align}

Acceleration modifies every matrix element through $q$. The off-diagonal entry $\alpha_L\sin2\theta$ measures detector coherence, while the diagonal entries $2\alpha_L\sin^2\theta$ and $2\alpha_L\cos^2\theta$ describe the population of the single-excitation states. As $a$ increases and $q\to1$, the coherence is suppressed, describing coherence suppression within the normalized effective-state model.

The corresponding correlation matrix is
\begin{equation}
T_L=\operatorname{diag}\left(
2\alpha_L\sin2\theta,\,
2\alpha_L\sin2\theta,\,
\gamma_L-2\alpha_L+\beta_L
\right).
\label{orig:32}\end{equation}
\subsection{Physical Limits and Model Applicability}
For a direct comparison of the two single-endpoint protocols, we also consider acceleration of detector $E$ while $D$ remains inertial. The transition channels are $|01\rangle\to|00\rangle$ and $|10\rangle\to|11\rangle$. Under the same first-order and normalization conventions as for detector $D$, the normalized final state, reduced density matrix, and correlation matrix are given in Appendix A.4, Eqs. \eqref{orig:33}--\eqref{orig:39}. Only these final results are used in the main-text comparison; the Dyson expansion, field trace, and Bogoliubov details for detector $E$ are given in Appendix A.4.

Comparing Eqs. \eqref{orig:34}--\eqref{orig:38} with the detector-$D$ expressions in Eqs. \eqref{orig:28}--\eqref{orig:31}, one sees that the initial-state asymmetry between the two detectors exchanges the roles of $\sin\theta$ and $\cos\theta$ when $E$ is accelerated. For the maximally entangled state $\theta=\pi/4$, the two protocols are exactly equivalent. The parameter $q$ remains the central Unruh variable; as $a$ increases, the coherence term $\alpha_R\sin2\theta$ is suppressed, describing the corresponding parameter dependence within the effective-state model.

Although the reduced-state matrices for acceleration of $D$ and $E$ have different forms, their $T^\dagger T$ spectra coincide for $\theta=\pi/4$. Consequently, for maximally entangled inputs the singular-value distribution of the network witness depends on the number of affected edges rather than on which endpoint is accelerated; for nonmaximal inputs the endpoint direction must be retained explicitly.

For $\theta=\pi/4$, define the dimensionless parameters
\begin{equation}
	u(q) = \frac{2(1-q)}{2(1-q) + \nu^2(1+q)},\quad \eta(q) = \frac{\nu^2(1+q)}{2(1-q)}.
\label{orig:40}\end{equation}

For the maximally entangled input, Eq. \eqref{orig:40} gives $T=\operatorname{diag}(u,u,u(\eta-1))$. The eigenvalues of $T^\dagger T$ are $\{u^2,u^2,u^2(\eta-1)^2\}$ and the singular values are $\{u,u,u|\eta-1|\}$. The two largest singular values must therefore be sorted as
\begin{equation}
(\tau_1,\tau_2)=\begin{cases}(u,u),&0\le\eta\le2,\\(u(\eta-1),u),&\eta>2.\end{cases}
\label{new:ordered}\end{equation}

For $0\le\eta\le2$, $u^2(\eta-1)^2\le u^2$; for $\eta>2$, the ordering reverses and $u(\eta-1)$ enters the leading pair. With $\nu^2=0.01$, the crossing occurs at $q_{\rm s}=399/401\simeq0.995012469$, well outside the controlled region $\eta\le0.1$. Appendix B lists the numerical values and verifies the derivative change on the two sides of the crossing.

At fixed $q$, the limit $\nu\to0$ removes the coupling correction and the density matrix approaches the initial pure state, with $\alpha_{L,R}\to1/2$ and $\beta_{L,R},\gamma_{L,R}\to0$. Thus both detector-$D$ and detector-$E$ acceleration recover $|\psi_\theta\rangle\langle\psi_\theta|$ in the zero-coupling limit.

As $q$ increases, field excitation and de-excitation leak population into the $|00\rangle$ and $|11\rangle$ sectors through $\gamma_{L,R}$ and $\beta_{L,R}$, while the coherence $\alpha_{L,R}\sin2\theta$ is suppressed. The limit $q\to1$ is only a formal extrapolation of the perturbative model. If the condition in Eq. \eqref{orig:20} fails near this limit, higher-order Dyson terms are required.

The first-order jump construction is used here as an effective resource state for the network calculation. It retains $U_1\rho_0U_1^\dagger$ but omits the same-order contributions $U_2\rho_0+\rho_0U_2^\dagger$, so normalization alone cannot establish a local CPTP map. Appendix \ref{app:trace} provides a quantitative test of this limitation. The strong-acceleration curves must therefore be interpreted as model extrapolations.

The effective state and its normalization do not certify complete positivity or trace preservation of the underlying dynamics. The spectral explanation of the local recovery at $\eta=2$ is explicit, but the physical meaning of the strong-acceleration continuation requires a higher-order derivation or a nonperturbative calculation. The extrapolated branch is not treated as evidence for a physical revival. The local recovery is not a result at any definite finite Dyson order: the normalization retains formal higher powers while the same-order $U_2\rho+\rho U_2^\dagger$ terms are omitted.

The local “V-shaped” recovery in the figures is an algebraic consequence of the spectral reordering and the associated switch of the optimal measurement direction. It occurs outside the perturbative regime and is not evidence that physical decoherence is suppressed or that nonlocality revives in the complete dynamics.

\subsection{Effective Edge and Network States for Network Calculations}\label{sec:networkstate}
For a seven-node tree, each independent source $S_e$ prepares $\rho_e^{(0)}=|\psi_{\theta_e}\rangle\langle\psi_{\theta_e}|$ on edge $e=(u,v)$. If one endpoint is accelerated, tracing out the field gives
\begin{equation}
	\rho_e=
	\begin{cases}
\rho_{\mathrm f}^{(L)}(\theta_e,q_e,\nu_e), & u \text{ accelerated},\\
\rho_{\mathrm f}^{(R)}(\theta_e,q_e,\nu_e), & v \text{ accelerated},\\
\rho_e^{(0)}, & u,v \text{ both inertial}.
	\end{cases}
\label{orig:41}\end{equation}
Here $q_e$ denotes the Unruh-response strength experienced by the accelerated detector on edge $e$.

Assuming independent edge fields and no field-mediated cross-correlations between distinct sources, the global pre-measurement state is
\begin{equation}
	\rho_{\text{net}} = \bigotimes_{e=1}^{M} \rho_e,
\label{orig:42}\end{equation}
Here $M$ is the number of sources. Equation \eqref{orig:42} is the state used in the network calculation. Tracing out the field independently on each edge produces acceleration-dependent mixing while preserving the source-independent tensor-product structure; nodes perform joint local measurements only at the final measurement stage.

For maximally entangled edge states ($\theta_e=\pi/4$), left- and right-endpoint acceleration give identical $T_e^\dagger T_e$ spectra, so the calculation can be compressed to the affected-edge number $r$. For nonmaximal inputs, Eq. \eqref{orig:41} must be evaluated separately for the accelerated endpoint.

\subsection{Network Environment Assumptions}
The tensor-product form in Eq. \eqref{orig:42} is a physical modeling assumption. Each edge is assigned an independent quantum field environment; detectors located at the same node are not assumed to share a field; and field-induced cross-correlations between distinct edges are neglected. These assumptions are separate from source independence, which is a property of the hidden-variable probability model in Eq. \eqref{orig:21}. If several detectors at one node couple to a common field, or if distinct edges have nonzero field correlations, then $\rho_{\mathrm{net}}=\bigotimes_e\rho_e$ and the edgewise singular-value factorization, including the $r$-compression, may fail. In that case the full global reduced state must be computed before applying a network witness.

\section{Seven-Node Tree Network and Its Nonlocality Witness}

\subsection{Network Structure and Six-Locality Model}

\begin{figure}[htbp]
\centering
\includegraphics[width=.85\linewidth]{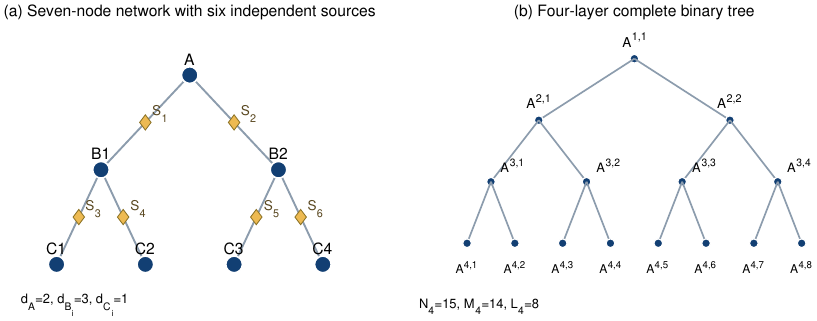}
\caption{Topology of the tree-structured quantum network. (a) Seven-node network with root $A$, branch nodes $B_1,B_2$, leaf nodes $C_i$, and six independent sources $S_i$. (b) Four-layer complete binary tree, where $A^{\ell,j}$ denotes the $j$th node at level $\ell$; $N_4=15$, $M_4=14$, and $L_4=8$.}

\label{fig:topology}
\end{figure}

Figure \ref{fig:topology} shows the seven-node network composed of root $A$, branch nodes $B_1,B_2$, and leaves $C_1,\ldots,C_4$. The six independent sources connect
\[
(A,B_1),\,(A,B_2),\,(B_1,C_1),\,(B_1,C_2),\,(B_2,C_3),\,(B_2,C_4).
\]
Let the six independent sources emit hidden variables $\lambda_1,\ldots,\lambda_6$. Under the network-local model, the joint distribution is
\begin{align}
	&P(o_A,b_1,b_2,c_1,c_2,c_3,c_4|x,y_1,y_2,z_1,z_2,z_3,z_4)\nonumber\\
	&=\int d\lambda_1\cdots d\lambda_6\,
	\prod_{e=1}^{6}P_e(\lambda_e)\,
	P(o_A|x,\lambda_1,\lambda_2)\,
	P(b_1|y_1,\lambda_1,\lambda_3,\lambda_4)\nonumber\\
	&\quad\times P(b_2|y_2,\lambda_2,\lambda_5,\lambda_6)
	\prod_{i=1}^{4}P(c_i|z_i,\lambda_{i+2}).
\label{orig:43}\end{align}

Each leaf $C_i$ is connected to source $\lambda_{i+2}$ and therefore depends only on that hidden variable.

To construct the network witness, define the following correlator combinations:
\begin{align}
	I_{\bm i,0}
	&=\frac1{16}\sum_{z_1,z_2,z_3,z_4}
	\left\langle A_{i_1}B^1_{i_2}B^2_{i_3}
	C^1_{z_1}C^2_{z_2}C^3_{z_3}C^4_{z_4}\right\rangle,\label{orig:44}\\
	I_{\bm j,1}
	&=\frac1{16}\sum_{z_1,z_2,z_3,z_4}
	(-1)^{z_1+z_2+z_3+z_4}
	\left\langle A_{j_1}B^1_{j_2}B^2_{j_3}
	C^1_{z_1}C^2_{z_2}C^3_{z_3}C^4_{z_4}\right\rangle,
\label{orig:45}\end{align}

Here $\bm i=(i_1,i_2,i_3)$ and $\bm j=(j_1,j_2,j_3)\in\{0,1\}^3$ are the measurement-input vectors for the root and two branch nodes. The observables $A_x$, $B_k^i$, and $C_z^i$ denote the corresponding measurements at nodes $A$, $B$, and $C$.

Yang et al. \cite{Yang2021} proved that every six-local correlation satisfies the following 64 inequalities, indexed by $i,j\in\{0,1\}^3$:

\begin{equation}
\boxed{
\left|I_{\bm i,0}\right|^{1/4}
+\left|I_{\bm j,1}\right|^{1/4}\leq1,
\qquad \bm i,\bm j\in\{0,1\}^{3}.
}
\label{orig:46}\end{equation}
Violation of any one of Eq. \eqref{orig:46} rules out the six-local model. Define the witness
\begin{equation}
\calS_6=
\left|I_{\bm i,0}\right|^{1/4}
 +\left|I_{\bm j,1}\right|^{1/4}.
\label{orig:47}\end{equation}
The condition $\calS_6>1$ is a sufficient witness of network nonlocality.

\subsection{Witness Values and Measurement Construction for Product Measurements}

For each edge state $\rho_e$, let $T_e$ be its correlation matrix and let $\tau_{e,1},\tau_{e,2}$ be the two largest singular values of $T_e^\dagger T_e$. Choose the corresponding orthogonal singular directions and define local Pauli observables $Z_{v,e}$ and $X_{v,e}$. These denote selected Pauli directions rather than fixed laboratory axes; directions may be chosen independently on different edges. The root measurements are
\begin{equation}
A_0=Z_{A,1}\otimes Z_{A,2},\qquad A_1=X_{A,1}\otimes X_{A,2},
\label{new:rootmeasurement}\end{equation}
At each branch node, use the product of the corresponding $Z$ observables and the product of the corresponding $X$ observables. At leaf $\ell$, choose
\begin{equation}
C_0^\ell=\cos\varphi_\ell Z_\ell+\sin\varphi_\ell X_\ell,\qquad
C_1^\ell=\cos\varphi_\ell Z_\ell-\sin\varphi_\ell X_\ell.
\label{new:leafmeasurement}\end{equation}
The factorized correlators are
\begin{align}
|I_{\bm0,0}|&=\left(\prod_{e=1}^6\tau_{e,1}\right)\prod_{\ell=1}^4|\cos\varphi_\ell|,\nonumber\\
|I_{\bm1,1}|&=\left(\prod_{e=1}^6\tau_{e,2}\right)\prod_{\ell=1}^4|\sin\varphi_\ell|.
\label{new:factor}\end{align}
Let $\mathfrak a=(\prod_e\tau_{e,1})^{1/4}$ and $\mathfrak b=(\prod_e\tau_{e,2})^{1/4}$. Hölder’s inequality gives $\calS_6\le[\mathfrak a\sum_\ell|\cos\varphi_\ell|+\mathfrak b\sum_\ell|\sin\varphi_\ell|]^{1/4}$; Cauchy--Schwarz then gives the upper bound in Eq. \eqref{orig:48}.
\begin{equation}
S_{6,\mathrm{prod}}^\star=\left[\left(\prod_{e=1}^6\tau_{e,1}\right)^{1/2}+\left(\prod_{e=1}^6\tau_{e,2}\right)^{1/2}\right]^{1/2}.
\label{orig:48}\end{equation}
Equation \eqref{orig:48} is the spectral-product witness used below. The superscript $\star$ denotes the value attainable within the specified product-measurement family; it is not a global optimum over arbitrary joint measurements. Related network witnesses may be optimized further \cite{Andreoli2017,Kundu2020}; the present construction is sufficient for the protocol comparison.

\section{Single- and Double-Node Acceleration in the Seven-Node \mbox{Network}}
This section analyzes how relativistic acceleration changes the product-measurement witness.

Consider the pure-state edge family
\begin{equation}
	|\psi_e\rangle = b_{e0}|00\rangle + b_{e1}|11\rangle.
\label{orig:49}\end{equation}

Define $b_e=2|b_{e0}b_{e1}|$, which is the concurrence of the pure edge state. The two largest singular values are $\tau_{e,1}=1$ and $\tau_{e,2}=b_e$, with the remaining squared values $1,b_e^2,b_e^2$.

Substitution into Eq. \eqref{orig:48} gives
\begin{equation}
	S_{6,\mathrm{prod}}^\star = \sqrt{1+\left(\prod_{e=1}^{6} b_e\right)^{1/2}}.
\label{orig:50}\end{equation}
Equation \eqref{orig:50} shows that any edge with nonzero concurrence can yield $S_{6,\mathrm{prod}}^\star>1$, so the network can violate the six-local inequality. For maximally entangled edges ($b_e=1$), the witness reaches $\sqrt2$.

\subsection{Parameter Conventions and Acceleration Protocols}

For reproducibility, we set all initial edges to $\theta=\pi/4$ and use $\nu^2=0.01$. We define node acceleration as applying the same UDW effective evolution to all detectors connected to the selected node, with at most one accelerated endpoint per edge. Let $r$ be the number of affected edges. For maximally entangled inputs, unaffected edges contribute the unit factor $1$, while the $r$ affected edges share the same singular-value distribution; the seven-node product-measurement value therefore reduces to
\begin{equation}
S_{6,\mathrm{prod}}^\star(r,q)=\left[\tau_1(q)^{r/2}+\tau_2(q)^{r/2}\right]^{1/2}.
\label{orig:52}\end{equation}

Here $\tau_1,\tau_2$ are the two largest singular values selected from $\{u,u,u|\eta-1|\}$ according to the ordering in Eq. \eqref{new:ordered}.

To probe the dependence on acceleration location, we select representative single- and double-node protocols. For single-node acceleration, leaf $C_1$, root $A$, and branch node $B_1$ correspond to $r=1,2,3$, respectively. For double-node acceleration, $(C_1,C_2)$, $(A,C_1)$, $(B_1,C_3)$, and $(B_1,B_2)$ correspond to $r=2,3,4,6$.

These values of $r$ count the affected edge resources directly. The reduction in Eq. \eqref{orig:52} applies only to maximally entangled inputs. For $\theta\ne\pi/4$, the full left- and right-endpoint states in Eq. \eqref{orig:28} must be evaluated edge by edge.

\subsection{Numerical Results and Interpretation}

\begin{figure}[htbp]
\centering\includegraphics[width=.94\linewidth]{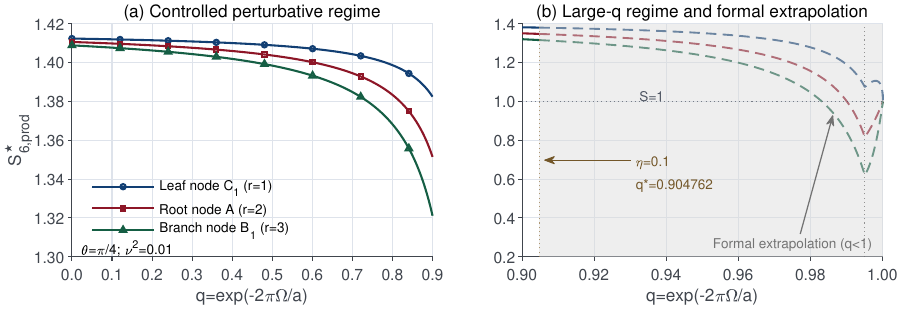}
\caption{Seven-node witness under single-node acceleration. All edges use $\theta=\pi/4$ and $\nu^2=0.01$; $r$ denotes the number of affected edges, with leaf, root, and branch-node acceleration corresponding to $r=1,2,3$. (a) $0\le q\le0.9$, the controlled perturbative regime; (b) $0.9<q<1$, the extrapolation regime, with solid curves for $\eta\le0.1$ and shaded/dashed curves for $\eta>0.1$. The boundary is $q_*=19/21$ and the dotted line $q_{\rm s}=399/401$ marks the spectral crossing; the shaded curves are algebraic extrapolations and are not physical-dynamics predictions.}
\label{fig:single}
\end{figure}
In Fig. \ref{fig:single}(a), the solid range satisfies $\eta\le0.1$. As $q$ increases, branch-node acceleration ($r=3$) decays fastest, followed by root acceleration ($r=2$), while leaf acceleration ($r=1$) is most robust. This confirms that the affected-edge number $r$ is the leading topological control parameter.

On the controlled branch, $\tau_1=\tau_2=u$ and $0\le\eta\le2$, so Eq. \eqref{orig:52} reduces to
\begin{equation}
	S_{6,\mathrm{prod}}^\star = \sqrt{2}\,u^{r/4}.
\label{orig:53}\end{equation}

\begin{figure}[htbp]
\centering\includegraphics[width=.92\linewidth]{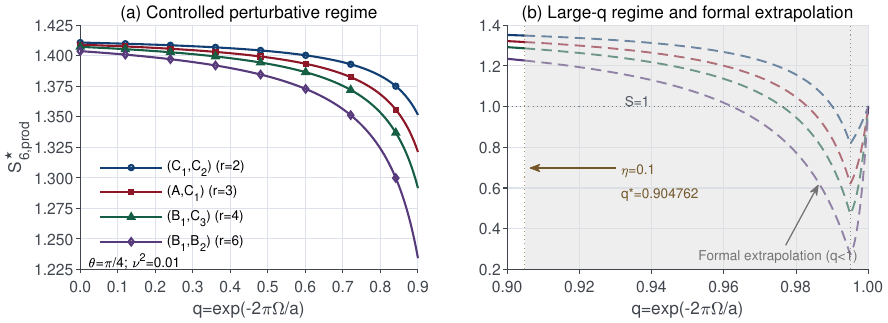}
\caption{Seven-node witness under nonadjacent double-node acceleration. The initial state is maximally entangled with $\theta=\pi/4$ and $\nu^2=0.01$. The four node pairs correspond to $r=2,3,4,6$. (a) $q\le0.9$ is the controlled perturbative regime; (b) $0.9<q<1$ is the extrapolation regime, with shading and dashed curves indicating $\eta>0.1$; these curves are algebraic extrapolations and are not physical-dynamics predictions.}
\label{fig:double}
\end{figure}

Equation \eqref{orig:53} shows that $r$ controls robustness on a fixed perturbative branch. The comparison in Fig. \ref{fig:double} follows the same ordering: the $r=6$ branch-node pair decays fastest, while simultaneous acceleration of two leaves with $r=2$ is most robust.

Figures 2 and 3 compare single-node and nonadjacent double-node acceleration. The witness curves decrease in the order predicted by $r$: $(B_1,B_2)$ with $r=6$, $(B_1,C_3)$ with $r=4$, $(A,C_1)$ with $r=3$, and $(C_1,C_2)$ with $r=2$. Protocols with the same $r$ coincide for maximally entangled inputs and identical environments, although node placement may matter for more general measurements.

In the controlled range $q\le0.9$, the witness decreases as acceleration increases. The branch node $B_1$ produces the largest loss, followed by the root $A$ and the leaf $C_1$, directly confirming Eq. \eqref{orig:53}.

The comparison between the two panels shows that protocols with the same affected-edge number can have identical witness curves. This coincidence is a consequence of maximal entanglement, identical edge environments, and the selected product-measurement family.

The right panels extend the normalized effective-state formula beyond $\eta=0.1$. The local recovery near $q\to1$ is caused by spectral reordering at $\eta=2$ and the accompanying measurement-direction switch; it is not evidence for physical revival. For $\nu^2=0.01$, the perturbative-control boundary $\eta=0.1$ corresponds to $q_*=19/21\simeq0.904762$, whereas the spectral crossing $\eta=2$ occurs at $q_{\rm s}=399/401\simeq0.995012$; the complete piecewise expression and numerical check are given in Appendix \ref{app:spectrum}.
\section{Extension to $n$-Layer Complete Binary Trees}

\subsection{Structure and Criterion}

We extend the seven-node tree to an $n$-layer ($n\ge2$) complete binary tree. The $i$th layer contains $2^{i-1}$ nodes, so the total number of nodes $N_n$, independent sources (edges) $M_n$, and leaves $L_n$ are
\begin{equation}
	N_n = 2^n - 1, \quad M_n = 2^n - 2, \quad L_n = 2^{n-1},
\label{orig:54}\end{equation}
The quantity $M_n$ is also the number of independent sources.

Assuming all sources are independent, consider the $(2^n-2)$-local hidden-variable model. As in the seven-node network, define correlator families $I_{\bm i,0}$ and $I_{\bm j,1}$ for internal-node input vectors $\bm i,\bm j$. The root holds two qubits, while every other internal node holds three, corresponding to degrees two and three. Source independence imposes the local bound
\begin{equation}
	|I_{\bm i,0}|^{1/L_n} + |I_{\bm j,1}|^{1/L_n} \le 1.
\label{orig:55}\end{equation}

The corresponding quantum witness is
\begin{equation}
S_{n,\mathrm{prod}}^\star=\left[\left(\prod_{e=1}^{M_n}\tau_{e,1}\right)^{2/L_n}+\left(\prod_{e=1}^{M_n}\tau_{e,2}\right)^{2/L_n}\right]^{1/2}.
\label{orig:56}\end{equation}

Equation \eqref{orig:56} naturally extends the product over $M_n$ edges and replaces the seven-node exponent $1/2$ by $2/L_n$, reflecting the constraint imposed by the leaves.

For $n=3$, $L_3=4$ and $M_3=6$, and Eq. \eqref{orig:56} reduces to the seven-node expression in Eq. \eqref{orig:48}, confirming consistency of the generalization.

\subsection{Depth Scaling and Werner Thresholds}

We define two acceleration protocols with explicit parameter conventions.

For affected edges with identical singular values and unaffected maximally entangled edges, Eq. \eqref{orig:56} gives
\begin{equation}
S_{n,\mathrm{prod}}^\star(r,q)=\left[\tau_1^{2r/L_n}+\tau_2^{2r/L_n}\right]^{1/2}
=\sqrt2(1+\eta)^{-r/L_n}\quad(\eta\le2).
\label{new:homogeneous}\end{equation}
Here $S_{n,\mathrm{prod}}^\star$ denotes the witness for an $n$-layer tree; the seven-node network is $n=3$. The direct scaling variable is $r/L_n$. The affected-edge fraction $r/M_n$ must be multiplied by $M_n/L_n$ to obtain the exponent in the witness.

\subsubsection{Protocol I: Root-Only Acceleration}
In this protocol, acceleration is applied only to the root node, so only two adjacent edges are affected by the relativistic effect, and $r=2$ does not change with the network depth $n$. The scaling factor $r/L_n$ in Eq. \eqref{orig:56} decreases as the depth increases; when the number of affected edges remains constant while the total number of leaves grows continuously with depth, the relative contribution of the same local perturbation to the product-measurement witness gradually decreases. This behavior is determined by the scaling relations of the network. Figure 4 shows the witness curves for root-only acceleration at different depths. Solid lines correspond to the controlled range and dashed lines to the formal extrapolation.

\begin{figure}[htbp]
\centering\includegraphics[width=.81\linewidth]{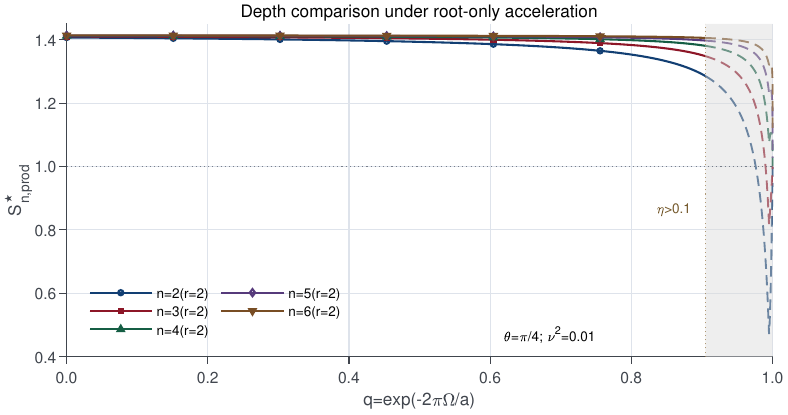}
\caption{Depth comparison for root-only acceleration. All edges start with $\theta=\pi/4$ and $\nu^2=0.01$, with $r=2$ affected edges and $n=2,\ldots,6$. The horizontal axis is $q$ and the vertical axis is $S_{n,\mathrm{prod}}^\star$. Solid curves denote $\eta\le0.1$; shading and dashed curves denote the extrapolation region $\eta>0.1$, with boundary $q_*=19/21$; the extrapolated curves are not physical-dynamics predictions.}
\label{fig:nroot}
\end{figure}
In the controlled region, the decrease of $S_{n,\mathrm{prod}}^\star$ becomes slower as $n$ increases: the decay is fastest at $n=2$, whereas a larger witness can be maintained at $n=6$ with $r=2$ fixed. This phenomenon originates from the scaling characteristics of the network: only the two edges connected to the root node are affected by the acceleration perturbation, whereas the total number of edges $M_n=2^n-2$ grows exponentially with depth.

The recovery effect appearing in the extrapolation region already lies beyond the range of applicability of perturbation theory, and a quantitative description of this region requires higher-order theory or nonperturbative dynamics.

\subsubsection{Protocol II: All Leaves Accelerated}
When all leaf nodes are accelerated simultaneously, the second endpoint of every leaf edge is subject to acceleration, so the number of perturbed edges is $r=L_n=2^{n-1}$. Substituting into Eq. \eqref{orig:56}, one sees that the factor $L_n$ exactly cancels the denominator $2/L_n$, and we finally obtain Eq. \eqref{orig:57}:
\begin{equation}
S_{n,\mathrm{prod}}^\star=\sqrt{\tau_1^2+\tau_2^2}.
\label{orig:57}\end{equation}

Equation \eqref{orig:57} shows that, for maximally entangled inputs and identical UDW parameters, the value of the witness under product measurements is independent of the network depth. Figure 5 displays the resulting curve collapse. Whatever the value of the depth $n$, the maximum of this witness remains constant. The formal recovery at large $q$ in Fig. 5 is a spectral-branch effect of the effective model; although the expression is continuous for $q<1$, it is not a complete finite-time dynamical prediction.

Comparing the two protocols shows that depth alone does not determine acceleration robustness; the core variable that directly controls this behavior is $r/L_n$, whereas the fraction $r/M_n$ must be topologically converted through $M_n/L_n$. In Protocol II, the noise distribution is amplified synchronously with the network size, thereby exhibiting depth invariance.

\subsubsection{Werner Noise and the White-Noise Threshold}

We now consider the complementary situation in which every edge is mixed with the same amount of white noise, rather than only a subset of edges being accelerated, and derive the visibility threshold that this perturbation imposes.

\begin{figure}[htbp]
\centering\includegraphics[width=.94\linewidth]{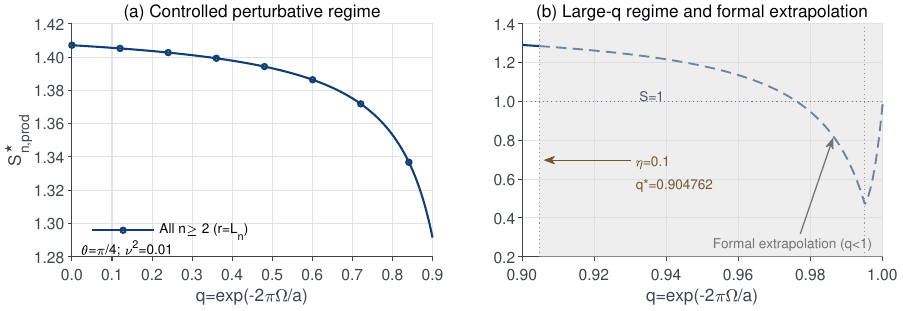}
\caption{Witness under simultaneous acceleration of all leaves. Each edge uses $\theta=\pi/4$, $\nu^2=0.01$, and $r=L_n$. (a) $q\le0.9$, where curves for $n\ge2$ coincide; (b) $0.9<q<1$, with solid curves for $\eta\le0.1$ and shaded/dashed curves for the formal extrapolation $\eta>0.1$; the extrapolated branch is not a physical-dynamics prediction.}
\label{fig:nleaf}
\end{figure}
Take each edge to be a Werner state with common visibility $0\le v\le1$, $\rho_e=v|\Phi^+\rangle\langle\Phi^+|+(1-v)\mathbb I/4$. Its correlation matrix is $T=v\,\mathrm{diag}(1,-1,1)$, whose singular values are $(v,v,v)$, so that $\tau_1=\tau_2=v$ on every edge. For the seven-node network the witness is
\begin{equation}
	S_{6,\mathrm{prod}}^\star = \sqrt{2}\,v^{3/2},\quad v > 2^{-1/3}.
\label{orig:51}\end{equation}
The initial maximally entangled state is $|\Psi^+\rangle$; since $|\Phi^+\rangle$ is locally Pauli-equivalent to it, the correlation spectrum and the witness value are unchanged. The condition $v>2^{-1/3}$ is nothing but the $n=3$ case of the general threshold derived below.

For an $n$-layer tree, substituting $\tau_1=\tau_2=v$ into Eq. \eqref{orig:56} together with $M_n/L_n=2-2^{2-n}$ gives
\begin{equation}
	S_{n,\mathrm{prod}}^\star = \sqrt{2}v^{M_n/L_n} = \sqrt{2}v^{2-2^{2-n}}.
\label{orig:58}\end{equation}

Equation \eqref{orig:58} reduces to Eq. \eqref{orig:51} at $n=3$. As $n$ increases, $M_n/L_n$ approaches 2 and the witness tends to $\sqrt2\,v^2$ in the deep-tree limit.

Setting $S_{n,\mathrm{prod}}^\star=1$, i.e.\ $v^{M_n/L_n}=2^{-1/2}$, yields the critical visibility
\begin{equation}
	v_{c,0}^{(n,\mathrm{prod})} = 2^{-L_n/(2M_n)} = 2^{-2^{n-1}/(2(2^n-2))}.
\label{orig:59}\end{equation}

Equation \eqref{orig:59} gives $v_{c,0}^{(n,\mathrm{prod})}\to2^{-1/4}\approx0.841$ as $n\to\infty$.

\begin{figure}[htbp]
\centering\includegraphics[width=.72\linewidth]{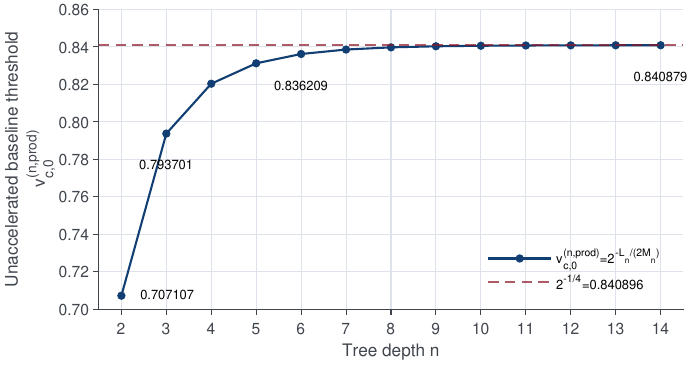}
\caption{Unaccelerated baseline visibility threshold for identical Werner sources. The horizontal axis is $n=2,\ldots,14$ and the vertical axis is $v_{c,0}^{(n,\mathrm{prod})}=2^{-L_n/(2M_n)}$. The dashed line marks the deep-tree limit $2^{-1/4}\simeq0.840896$; $v>v_{c,0}^{(n,\mathrm{prod})}$ guarantees violation of the selected source-independent local bound.}
\label{fig:vis}
\end{figure}

Figure \ref{fig:vis} shows this unaccelerated white-noise baseline: $v_{c,0}^{(n,\mathrm{prod})}$ rises from about $0.71$ at $n=2$ and is essentially saturated by $n=9$, since $L_n/(2M_n)=1/4+O(2^{-n})$ converges rapidly; the finite ceiling $2^{-1/4}$ therefore characterizes the selected witness rather than a fundamental boundary. Two qualifications apply. For the selected product-measurement witness and effective resource model, the threshold is a sufficient rather than a necessary condition: $v>v_{c,0}^{(n,\mathrm{prod})}$ guarantees violation of the corresponding $(2^n-2)$-local bound, whereas $v\le v_{c,0}^{(n,\mathrm{prod})}$ does not imply network locality. It is moreover an unaccelerated baseline, distinct from any threshold $v_c(n,q)$ for simultaneous acceleration and noise.

The present result has the same structure as the acceleration results above. In both cases the witness takes the form $\sqrt2\,f^{\,R/L_n}$, where $R$ is the number of degraded edges and $f$ the single-edge degradation factor: uniform noise degrades all $M_n$ edges with $f=v$, whereas localized acceleration degrades only $r$ edges with $f=(1+\eta)^{-1}$. Thus the exponent is set by how many edges are degraded rather than by what degrades them: the scaling form is dictated by the network topology, and the two perturbations differ only in $R$ and $f$.

\section{Conclusion, Scope, and Outlook}

Within a weak-coupling, first-order effective-resource model, this work has examined how relativistic acceleration modifies source-independent nonlocality witnesses in tree-structured quantum networks and has extended the seven-node analysis to complete binary trees of arbitrary depth. The conclusions below rely on independent edge environments, maximally entangled inputs, at most one accelerated endpoint per edge, and the product-measurement construction adopted at internal nodes.

The analysis leads to three closely related model-internal conclusions. Source independence factorizes the network witness into a product of edge resources. With orthogonal singular directions and the internal-node product measurements used here, the product-measurement values for both the seven-node network and the general complete binary tree are determined by the two largest singular values of each edge correlation matrix, which separates the local detector dynamics from the network criterion and allows different edge states and acceleration protocols to be compared within one framework. For maximally entangled inputs in the controlled perturbative regime, the leading acceleration dependence is determined by the number $r$ of affected edges. Thus, for fixed node types, acceleration parameters, and edge states, a higher-degree node modifies more edge resources and produces a faster witness decay, subject to the measurement strategy and perturbative branch specified here. For an $n$-layer complete binary tree, the direct dimensionless scaling variable is $r/L_n$: under root-only acceleration, $r=2$ remains fixed while $L_n=2^{n-1}$ grows with depth, so the acceleration-induced suppression is relatively diluted on the controlled branch, whereas simultaneous acceleration of all leaves gives $r=L_n$ and a depth-independent witness. For identical Werner visibility, the critical threshold of the selected witness approaches $2^{-1/4}$ as the depth increases. More generally, the witness retains the form $\sqrt2\,f^{\,R/L_n}$ with $R$ the number of degraded edges and $f$ the single-edge degradation factor, so the exponent is fixed by the network topology rather than by the degradation mechanism.

This paper describes a scaling behavior of the following kind: when the number of perturbed edges remains constant while the number of leaf nodes grows with the network depth, the influence of the acceleration perturbation is relatively weakened. This conclusion is a scaling relation obtained in the present work, rather than a citation of an existing standard term. In the extrapolated results for the strong-acceleration region, the ordering of the singular values changes at $\eta=2$, and the witness curve may show a local recovery. This recovery originates from a switch between the two spectral branches of the effective correlation matrix, and cannot be understood as a physical revival of nonlocality under the complete dynamics.

The conclusions have several limitations. First, the effective states constructed in Sec.~III and Appendix A.4 are obtained by first-order amplitude truncation followed by renormalization. Normalization and positive semidefiniteness alone do not establish that this state arises from a complete local CPTP dynamics; the trace-preservation check is given in Appendix \ref{app:trace}. Once $\eta$ approaches or exceeds the perturbative domain, same-order trace-preserving terms must be retained and tested using a higher-order Dyson expansion or a master equation. Second, the $r$-compression applies only to maximally entangled inputs with at most one accelerated endpoint per edge. Asymmetric inputs or multiple detectors coupled to a common field require explicit left/right endpoint labels and possible cross-correlations. Finally, $\calS>1$ is only a sufficient condition for the selected measurement strategy, not a necessary condition for network nonlocality; failure to reach this threshold does not imply network locality.

Future work should develop higher-order UDW dynamics to determine the validity boundary of the extrapolated strong-acceleration branch, analyze cross-correlations and non-Markovian effects when multiple detectors at one node couple to a common field, and extend the framework to arbitrary branching numbers and irregular trees. Optimized measurements or semidefinite-programming methods may then provide witnesses stronger than the product-measurement criterion used here.

\clearpage\appendix
\renewcommand{\thesection}{\Alph{section}}
\renewcommand{\thesubsection}{\Alph{section}.\arabic{subsection}}
\renewcommand{\thesubsubsection}{\Roman{subsubsection}}
\numberwithin{equation}{section}
\section{Details of the Detector--Field Expansion}\label{app:qft}
\subsection{Spatial Smearing and the Interaction Picture}
In the interaction picture, the UDW interaction Hamiltonian along detector proper time $\tau$ is
\begin{equation}
	H_I(\tau) = \sum_{i=D,E} g_i(\tau) \left[ D_i e^{-i\Omega \tau} + D_i^\dagger e^{i\Omega \tau} \right] \Phi_{\chi_i}(\tau),
\label{orig:7}\end{equation}
Here $D_D=D$ and $D_E=E$; $g_i(\tau)$ are real switching functions controlling the interaction strength and duration, and $\Phi_{\chi_i}(\tau)$ is the smeared field operator:
\begin{equation}
	\Phi_{\chi_i}(\tau) = \int_{\Sigma_\tau} d^3\xi \sqrt{h}\, \chi_i(\xi)\phi\big(x_i(\tau,\xi)\big).
\label{orig:8}\end{equation}
The hypersurface $\Sigma_\tau$ is the constant-proper-time hypersurface, $h$ is the determinant of its induced metric, and $\chi_i(\xi)$ is the detector spatial profile.

We use a Gaussian spatial profile
\begin{equation*}
	\chi(\boldsymbol{\xi})
	=
	\frac{1}{(\kappa\sqrt{2\pi})^3}
	\exp\!\left(
	-\frac{|\boldsymbol{\xi}|^2}{2\kappa^2}
	\right),
\end{equation*}

Here $\kappa$ is the effective detector size. The profile decays rapidly away from the detector; finite-size effects can be retained through the smearing function.

For interaction duration $\Delta$, the single-detector Hamiltonian becomes
\begin{equation*}
	\begin{aligned}
		H_I(\tau) &= \lambda_{\mathrm c}g(\tau) &\left[ e^{-\mathrm{i}\Omega\tau} D \otimes \Phi_\chi(\tau) + e^{\mathrm{i}\Omega\tau} D^\dagger \otimes \Phi_\chi^\dagger(\tau) \right].
	\end{aligned}\end{equation*}

For the single-endpoint acceleration of detector $D$, we set $D_i=D$ and denote the coupling constant by $\lambda_{\mathrm c}$, distinct from the associated test-function parameter $\lambda_{\mathrm f}$; the switching function $g(\tau)$ is supported on $[\tau_0,\tau_0+\Delta]$. The phase factors arise from $D(\tau)=D e^{-i\Omega\tau}$ under $H_D=\Omega D^\dagger D$, while $\Phi_\chi(\tau)$ is the spatially smeared field.
\subsection{Dyson Expansion and the Joint State}
\begin{equation}
	U_I
	=
	\mathcal{T}
	\exp\!\left[
	-\mathrm{i}
	\int_{\tau_0}^{\tau_0+\Delta}
	H_I(\tau)\,
	\mathrm{d}\tau
	\right],
\label{orig:10}\end{equation}

Here $\mathcal T$ denotes proper-time ordering.
For weak coupling ($\lambda_{\mathrm c}\ll1$), the Dyson expansion is
\begin{equation}
	U_I \approx \mathbb{I} - \mathrm{i}\int_{\tau_0}^{\tau_0+\Delta} H_I(\tau)\mathrm{d}\tau + O(\lambda_{\mathrm c}^2).
\label{orig:11}\end{equation}

Substitution of $H_I(\tau)$ into Eq. \eqref{orig:11} gives
\begin{equation}
	U^{(1)} = \mathbb{I} - \mathrm{i}\lambda_{\mathrm c}
	\left[
	D \int_{\tau_0}^{\tau_0+\Delta} \mathrm{d}\tau\, g(\tau) e^{-\mathrm{i}\Omega\tau} \Phi_\chi(\tau)
	+ D^\dagger \int_{\tau_0}^{\tau_0+\Delta} \mathrm{d}\tau\, g(\tau) e^{\mathrm{i}\Omega\tau} \Phi_\chi^\dagger(\tau)
	\right].
\label{orig:12}\end{equation}
For compactness, define the effective smearing integrals
\begin{equation}
\begin{aligned}
\phi(f)&=\int_{\tau_0}^{\tau_0+\Delta}d\tau\,g(\tau)e^{-i\Omega\tau}\Phi_\chi(\tau),\\
\phi^\dagger(f)&=\int_{\tau_0}^{\tau_0+\Delta}d\tau\,g(\tau)e^{i\Omega\tau}\Phi_\chi^\dagger(\tau).
\end{aligned}
\label{orig:13}\end{equation}

Substituting Eq. \eqref{orig:13} yields the compact first-order operator
\begin{equation*}
	U^{(1)} = \mathbb{I} - \mathrm{i}\lambda_{\mathrm c} \big[ D\phi(f) + D^\dagger \phi^\dagger(f) \big].\end{equation*}
Here $f$ contains the switching function, transition phase, and spatial profile; Eq. \eqref{orig:13} keeps temporal switching and spatial smearing distinct.
The detector--field state after the interaction is
\begin{equation}
	\ket{\Psi_{\mathrm f}^{DE\phi}}
	=
	U^{(1)}
	\ket{\Psi_{\tau_0}^{DE\phi}}
	+
	O(\lambda_{\mathrm c}^2).
\label{orig:15}\end{equation}

The positive-frequency term $D$ describes detector de-excitation and the negative-frequency term $D^\dagger$ detector excitation. Both processes change the field state and create detector--field correlations; in this paper their effect is retained only as the coherence and population correction of the effective resource state.

\subsection{Bogoliubov Transformation and Field Excitation}
The accelerated detector worldline lies in Rindler wedge I.

The corresponding modes are described in wedge I. A Bogoliubov transformation connects Rindler and Minkowski modes.

For the mode of frequency $\Omega$, the transformation is
\begin{equation*}
	a_{\mathrm{RI}}(\bar{\lambda_{\mathrm f}})
	=
	\frac{
		a_M(F_{1\Omega})
		+
		\mathrm{e}^{-\pi\Omega/a}
		a_M^\dagger(F_{2\Omega})
	}{
		\sqrt{1-\mathrm{e}^{-2\pi\Omega/a}}
	},\end{equation*}

\begin{equation}
	a_{\mathrm{RI}}^\dagger(\lambda_{\mathrm f})
	=
	\frac{
		a_M^\dagger(F_{1\Omega})
		+
		\mathrm{e}^{-\pi\Omega/a}
		a_M(F_{2\Omega})
	}{
		\sqrt{1-\mathrm{e}^{-2\pi\Omega/a}}
	}.
\label{orig:17}\end{equation}
Here $\lambda_{\mathrm f}=-\lambda_{\mathrm c}\mathcal P_+\mathcal Gf$, with $\mathcal P_+$ the positive-frequency projection and $\mathcal G$ the causal propagator. $F_{1\Omega}$ and $F_{2\Omega}$ are coupling-selected wave packets whose explicit forms follow from the Minkowski test function and wedge reflection.
\[
F_{1\Omega} = \frac{\lambda_{\mathrm f} + e^{-\pi\Omega/a}\lambda_{\mathrm f} \circ w}{\left(1 - e^{-2\pi\Omega/a}\right)^{1/2}},\quad
F_{2\Omega} = \frac{\overline{\lambda_{\mathrm f} \circ w} + e^{-\pi\Omega/a}\overline{\lambda_{\mathrm f}}}{\left(1 - e^{-2\pi\Omega/a}\right)^{1/2}},
\]

The map $w(t,x,y,z)=(-t,-x,y,z)$ sends wedge I to wedge II; $\lambda_{\mathrm f}\circ w$ denotes composition and $\overline{\lambda_{\mathrm f}}$ complex conjugation. The transformation implies the Unruh temperature $T_U=a/(2\pi)$ and the thermal weight used in the effective resource state; the finite-time response is given by Eq. \eqref{new:finite_response}.
Using
\begin{equation}
	a_M(F_{i\Omega})\ket{0_M}=0,
\label{orig:18}\end{equation}

one obtains the corresponding field-excitation and detector-transition amplitudes.

\subsection{Acceleration of Detector $E$}

When $E$ is accelerated while $D$ remains inertial, the transition channels are $|01\rangle\to|00\rangle$ and $|10\rangle\to|11\rangle$. Applying the same Dyson, Bogoliubov, and field-trace framework gives, to first order,
\begin{equation}
	\ket{\Psi_{\mathrm f}^{DE\phi}}
	=\ket{\Psi_{\tau_0}^{DE\phi}}
	+\frac{\nu}{\sqrt{1-q}}\left[\sin\theta\,\ket{00}\ket{1_{\widetilde F_1}}
	+e^{-\pi\Omega/a}\cos\theta\,\ket{11}\ket{1_{\widetilde F_2}}\right],
\label{orig:33}\end{equation}
The normalization factor is
\begin{equation}
	\mathcal{N}_R = 1 + \frac{\nu^2\left(\sin^2\theta + q\cos^2\theta\right)}{1 - q}.
\label{orig:34}\end{equation}

Thus the final reduced detector state for acceleration of $E$ is
\begin{equation}
\rho_{\mathrm f}^{(R)}=
\begin{pmatrix}
\gamma_R&0&0&0\\
0&2\alpha_R\sin^2\theta&\alpha_R\sin2\theta&0\\
0&\alpha_R\sin2\theta&2\alpha_R\cos^2\theta&0\\
0&0&0&\beta_R
\end{pmatrix},
\label{orig:35}\end{equation}
where
\begin{align}
\alpha_R&=\frac{1-q}{2(1-q)+2\nu^2(\sin^2\theta+q\cos^2\theta)},\label{orig:36}\\
\beta_R&=\frac{\nu^2q\cos^2\theta}{(1-q)+\nu^2(\sin^2\theta+q\cos^2\theta)},\label{orig:37}\\
\gamma_R&=\frac{\nu^2\sin^2\theta}{(1-q)+\nu^2(\sin^2\theta+q\cos^2\theta)}.
\label{orig:38}\end{align}
Accordingly,
\begin{equation}
T_R=\operatorname{diag}\left(
2\alpha_R\sin2\theta,\,
2\alpha_R\sin2\theta,\,
\gamma_R-2\alpha_R+\beta_R
\right).
\label{orig:39}\end{equation}

\section{Numerical Check of the Spectral Crossing and Extrapolated Curves}\label{app:spectrum}
Using $\eta=\nu^2(1+q)/[2(1-q)]$, we obtain
\begin{equation}
q(\eta)=\frac{2\eta-\nu^2}{2\eta+\nu^2},\qquad
q_*=\frac{0.2-\nu^2}{0.2+\nu^2},\qquad q_{\rm s}=\frac{4-\nu^2}{4+\nu^2}.
\label{app:qboundary}\end{equation}
The value $q_*$ corresponds to $\eta=0.1$, while $q_{\rm s}$ corresponds to the crossing $\eta=2$. For $\nu^2=0.01$, the numerical values are listed in Table~\ref{tab:spectrum}.
\begin{table}[htbp]\centering
\caption{Singular values and ordering across the spectral crossing. Here $u=(1+\eta)^{-1}$ and $z=u|\eta-1|$; $q<1$ and values are rounded to six decimal places.}
\begin{tabular}{ccccc}\toprule
$\eta$&$q$&$u$&$z$&$(\tau_1,\tau_2)$\\\midrule
0.1&0.904762&0.909091&0.818182&$(u,u)$\\
1.9&0.994751&0.344828&0.310345&$(u,u)$\\
2.0&0.995012&0.333333&0.333333&$(u,u)$\\
2.1&0.995249&0.322581&0.354839&$(z,u)$\\\bottomrule
\end{tabular}\label{tab:spectrum}\end{table}
For $p=r/L_n>0$, the sorted singular values yield the complete algebraic branch
\begin{equation}
S_{n,\mathrm{prod}}^\star(\eta)=\begin{cases}
\sqrt2(1+\eta)^{-p},&\eta\le2,\\
(1+\eta)^{-p}\sqrt{(\eta-1)^{2p}+1},&\eta>2.
\end{cases}\label{app:piecewise}\end{equation}
The function is continuous at $\eta=2$, with value $S_{\rm s}=\sqrt2\,3^{-p}$ and one-sided derivatives
\begin{equation}
\left.\frac{dS}{d\eta}\right|_{2^-}=-\frac p3S_{\rm s},\qquad
\left.\frac{dS}{d\eta}\right|_{2^+}=\frac p6S_{\rm s}.
\label{app:derivatives}\end{equation}
The crossing is therefore a local minimum and produces the V-shaped turn. The derivative change is caused by switching the singular direction entering the product-measurement witness, not by recovery of $u$, which decreases monotonically with $\eta$. The second branch need not increase for every $p$ over its full domain, so it cannot be summarized as stronger acceleration always enhancing nonlocality. This is an algebraic property of the effective model, not a physical interpretation of the extrapolated recovery.

\section{Trace-Preservation Check for the Effective State}\label{app:trace}
If an unconditional local CPTP channel acts only on $D$, the marginal of the untouched detector $E$ should remain unchanged. The effective state in Eq. \eqref{orig:28}, however, gives
\begin{equation}
\widetilde P_E(0)=\frac{\cos^2\theta(1+p_\downarrow)}{1+p_\downarrow\cos^2\theta+p_\uparrow\sin^2\theta},\qquad
p_\downarrow=\frac{\nu^2}{1-q},\quad p_\uparrow=\frac{\nu^2q}{1-q}.
\label{app:marginal}\end{equation}
The result generally differs from the initial $\cos^2\theta$ because the density matrix built from $U^{(1)}=\Id+U_1$ contains $U_1\rho U_1^\dagger$, while a consistent second-order expansion must also retain
\begin{equation}
\rho_f=\rho+U_1\rho+\rho U_1^\dagger+U_1\rho U_1^\dagger+U_2\rho+\rho U_2^\dagger+O(\lambda_{\mathrm c}^3).
\label{app:secondorder}\end{equation}
Global normalization cannot restore the missing same-order terms. We therefore use the result only as an effective resource model, not as a complete unconditional local channel. A complete treatment must restore all same-order terms and check trace preservation, complete positivity, and approximation error.

\end{document}